\documentclass[aps,preprint]{revtex4}%
\usepackage{amsfonts}
\usepackage{amsmath}
\usepackage{amssymb}
\usepackage{graphicx}%
\begin{document}
\title{Fading Hawking Radiation }
\author{I.Sakalli}
\email{izzet.sakalli@emu.edu.tr}
\author{M. Halilsoy}
\email{mustafa.halilsoy@emu.edu.tr}
\author{H.Pasaoglu}
\email{hale.pasaoglu@emu.edu.tr}
\affiliation{Department of Physics, Eastern Mediterranean University,G. Magusa, North
Cyprus, Mersin-10, Turkey.}

\begin{abstract}
In this study, we explore a particular type Hawking radiation which ends with
zero temperature and entropy. The appropriate black holes for this purpose are
the linear dilaton black holes. In addition to the black hole choice, a recent
formalism in which the Parikh-Wilczek's tunneling formalism amalgamated with
quantum corrections to all orders in $\hbar$ is considered. The adjustment of
the coefficients of the quantum corrections plays a crucial role on this
particular Hawking radiation. The obtained tunneling rate indicates that the
radiation is not pure thermal anymore, and hence correlations of outgoing
quanta are capable of carrying away information encoded within them. Finally,
we show in detail that when the linear dilaton black hole completely
evaporates through such a particular radiation, entropy of the radiation
becomes identical with the entropy of the black hole, which corresponds to "no
information loss".

\end{abstract}
\maketitle

\section{INTRODUCTION}

Hawking \cite{hawking74,hawking75,hawking76} and \cite{bekenstein73} showed in
their seminal works that a black hole (BH) should slowly radiate away energy
with its characteristic temperature and entropy. But the semi-classical
picture of the Hawking radiation has a thermal nature, which poses a
fundamental physical problem. Because, when the material entering the BH is a
pure quantum state, the transformation of that state into the mixed state of
Hawking radiation would destroy information about the original quantum state.
However, this violates quantum mechanical unitarity and presents a physical
paradox -- so called the information loss paradox. For review of the topic and
references on the BH information loss problem the reader may refer to
\cite{preskill92,page94,russo05}. There are various ideas about how the
paradox could be solved. Among them, may be the most elegant and
comprehensible one is the Parikh and Wilczek (PW)'s quantum tunneling
formalism \cite{parikh00}. Their tunneling formalism is based on the null
geodesics together with the WKB method. They showed explicitly how the
inclusion of back-reaction effects, which guarantees the conservation of
energy during a particle tunneling the horizon, yields a non-thermal
correction to the BH radiation spectrum. For a recent review of
\textquotedblleft tunneling methods and Hawking radiation\textquotedblright%
\ one may consult \cite{vanzo11}. On the other hand, the form of their
non-thermal correction had a shortcoming since they did not consider the
Planck-scale ($\hbar$) quantum corrections, which elicit correlations between
quanta emitted with different energies. The first attempt to fix this
shortcoming came from \cite{arzano05}, who proposed a modified version of the
tunneling picture in which a leading order Planck-scale quantum correction was
introduced. In addition to this, \cite{banerjee08} have recently provided a
general framework for studying quantum corrections to all orders in $\hbar$ to
the entropy of a BH. When the effects of the quantum corrections are
neglected, one recovers the PW's results of the BH \cite{parikh00}. Although
there are supportive studies, see for
instance\cite{majhi09,banerjee09,zhu09,akbar10,mirza11,sheyki12}, to
\cite{banerjee08}, in recent times their work has been under criticism by
\cite{yale11}, who claimed that Banerjee and Majhi's result assumes an
incorrect definition of energy. Putting aside these discussions, here we
concentrate on the study which has recently been published by Singleton,
Vagenas, Zhu and Ren (SVZR) \cite{singleton10}. They have attempted to show
that the quantum corrections to all orders in $\hbar$ can be adjusted finely
so much so that both entropy and temperature of the Schwarzschild BH go to
zero as the mass of the BH is radiated away, i.e. $S,T(M\rightarrow
0)\rightarrow0$. But, immediately after it is understood that such a scenario
is not possible for the Schwarzschild BH \cite{singleton11}. In fact, the key
idea of the present study is to examine whether the quantum corrected Hawking
radiation with $S,T(M\rightarrow0)\rightarrow0$ is possible for other types of
BHs or not. In this article, we consider a general class of 4-dimensional
($4D$) metric which belongs to static, spherically symmetric linear dilaton
black holes (LDBHs) \cite{clement03,clement07} that constitute solutions to
Einstein-Maxwell-Dilaton (EMD), Einstein-Yang-Mills-Dilaton (EYMD) and
Einstein-Yang-Mills-Born-Infeld-Dilaton (EYMBID) theories \cite{mazhari09}.
The LDBHs are known to be a special class of non-asymptotically flat (NAF)
spacetimes. The reason why we focus on the LDBHs is that by using merely the
PW's quantum tunneling formalism one can not modify their thermal character of
the Hawking radiation \cite{pasa09}. This means that the original PW's
tunnelingformalism fails to answer the information loss paradox appearing in
the LDBHs. Because of this, in addition to the back reaction effects we need
to take into account the quantum corrections to obtain a radiation other than
pure thermal \cite{sakalli11}. As an extension of the study \cite{sakalli11},
here we consider the general form of the quantum corrected temperature given
by SVZR, and apply it to the LDBHs in order to derive specific entropy and
temperature, both of which go to zero with $S,T(M\rightarrow0)\rightarrow0$.
Detailed calculations of these processes are given in the next sections, and
as a result we obtain the above-mentioned radiation, and it is not pure
thermal. The behaviors of both the entropy and temperature of the LDBH with
the quantum correction parameters coming from String Theory (ST) and Loop
Quantum Gravity (LQG) are examined. We find that the results which have no any
physical ambiguity are possible only in the LQG case. Moreover, it is
highlighted that higher order quantum corrections which are in conform with
the back reaction effects provide the correlations between the emitted quanta.
Finally, we show that the LDBHs are able to evaporate away completely with the
entropy conservation (initial BH entropy is equal to the entropy of the
radiation), which leads to the fact that information is not lost. Organization
of the paper is as follows. In Sect. 2, we derive the entropy and temperature
providing $S,T(M\rightarrow0)\rightarrow0$ in quantum corrected LDBHs. Sect. 3
is devoted to the entropy conservation argument and Sect. 4 completes the
paper with discussion and conclusion. 

Throughout the paper, the units $G=c=k_{B}=1$ and $L_{p}^{2}=\hbar$ are used.

\section{QUANTUM CORRECTED ENTROPY AND TEMPERATURE EXPRESSIONS FOR $4D$-LDBHs}

As it was shown in \cite{mazhari09}, $4D$-LDBHs in EMD, EYMD and EYMBID
theories are described by the metric%

\begin{equation}
ds^{2}=-fdt^{2}+\frac{dr^{2}}{f}+R^{2}d\Omega_{2}^{2},
\end{equation}

with the metric functions%

\begin{equation}
f=\tilde{\Sigma}(r-r_{+}),\text{ \ }R=A\sqrt{r},
\end{equation}

It is obvious that metric (1) represents a static, non-rotating BH with a
horizon at $r_{+}.$ The dimensional constants $\tilde{\Sigma}$ and $A$ in the
metric functions (2) take different values according to the concerned theory
(EMD, EYMD or EYMBID) \cite{mazhari09}. For $r_{+}\neq0$, the horizon hides
the naked singularity at $r=0$. However, in the extreme case of $r_{+}=0,$ the
central null singularity at $r=0$ is marginally trapped in which it does not
allow outgoing signals to reach external observers. Namely, even in the
extreme case of $r_{+}=0,$ metric (1) maintains its BH property.

By using the definition of quasi-local mass $M$ \cite{brown93} for the NAF
metric (1), one finds a relation between the horizon $r_{+}$\ and the mass $M$
as
\begin{equation}
r_{+}=\frac{4M}{\tilde{\Sigma}A^{2}}.
\end{equation}

After some elementary dimensional analysis, one can see that the units of $M$
and $A^{2}$ are $L_{p}$, while $\tilde{\Sigma}$ has the unit of $L_{p}^{-1}$
so that $r_{+}$ has the unit of $L_{p}$.

Recently, it has been shown that the temperature for a general class of
static, spherically symmetric BH with quantum corrections to all orders in
$\hbar$\ \cite{singleton10} is given by%

\begin{equation}
T=\frac{\hbar\kappa}{2\pi}\left(  1+\sum_{j=1}^{\infty}\frac{\alpha_{j}%
\hbar^{j}}{r_{+}^{2j}}\right)  ^{-1},
\end{equation}

where $\kappa$\ is the surface gravity of the BH such that it becomes
$\kappa=\frac{\tilde{\Sigma}}{2}$ for the LDBHs, and $\alpha_{j}$'s --
dimensionless constants -- stand for the quantum correction terms. In this
expression $\frac{\hbar\kappa}{2\pi}$ is nothing but the well-known Hawking
temperature $T_{H}$. Here, we wish to highlight one of the important features
of the LDBHs that the Hawking temperature of the LDBH, $T_{H}=\frac
{\hbar\tilde{\Sigma}}{4\pi},$ is independent of their quasi-local mass $M$,
and which is therefore a constant throughout the evaporation process i.e. an
isothermal process.

In general, the first law of thermodynamics is about an expression for the
entropy ($S$) as%

\begin{equation}
S=\int\frac{dM}{T},
\end{equation}

where $M$ is the total energy (mass) of the BH. As we adopt the temperature
with generic quantum corrections from (4), the entropy to all orders in
$\hbar$ can be found by substituting (4) into (5), and by evaluating the
integral. Thus, for the LDBHs one obtains the following modified entropy as a
function of $M$%

\begin{equation}
S(M)=\frac{M}{T_{H}}\left(  1-\sum_{j=1}^{\infty}\frac{\alpha_{j}}{2j-1}%
x^{j}\right)  .
\end{equation}

\bigskip where $x=\frac{\hbar\tilde{\Sigma}^{2}A^{4}}{16M^{2}}$ is a
dimensionless quantity.

As mentioned before, our ultimate aim is to find a specific condition by which
it leads to a complete radiation of the LDBH with $S,T(M\rightarrow0)$
$\rightarrow0$. This requirement implies conditions on the $\alpha_{j}$'s. It
is remarkable to see that the only possibility which satisfies
$S,T(M\rightarrow0)$ $\rightarrow0$ is,%

\begin{equation}
\alpha_{j}=\frac{(-1)^{j+1}(2j-1)}{j}\alpha_{1},
\end{equation}

Inserting this into the sum of (6), we find the modified LDBH entropy as%

\begin{equation}
S(M)=\frac{M}{T_{H}}\left[  1+\alpha_{1}\ln(\frac{16M^{2}}{16M^{2}+\hbar
\tilde{\Sigma}^{2}A^{4}})\right]  ,
\end{equation}

Now, it can be easily checked that $S(M\rightarrow0)\rightarrow0$\ and
$S(M\rightarrow\infty)\rightarrow\infty$. Although the result of the sum in
(8) stipulates that $M>\frac{\sqrt{\hbar}\tilde{\Sigma}A^{2}}{4}$, by making
an analytical extension of the zeta function \cite{singleton10,abram65}, one
can redefine the sum via $\alpha_{1}\ln(\frac{16M^{2}}{16M^{2}+\hbar
\tilde{\Sigma}^{2}A^{4}})$ such that it becomes valid also for $M<\frac
{\sqrt{\hbar}\tilde{\Sigma}A^{2}}{4}$. We plot $S(M)$ (8) versus $M$ for the
cases of semi-classical and quantum corrections to all orders in $\hbar$, and
display all graphs in Fig. 1. In all figures, we have used two different
$\alpha_{1}$\ values such that $\alpha_{1}=-\frac{1}{2}$ is taken as the
representative of the LQG \cite{meissner04}, while the choice $\alpha
_{1}=\frac{1}{2}$ stands for the ST \cite{solodukhin98,zwiebach04}. Here,
physically inadmissible case belongs to the ST's one in which the behavior of
the entropy is not well-defined. Because, as seen in Fig. 1(b), just before
the complete evaporation of the LDBH, the entropy first decreases to a
negative value and then increases from below to become zero with $M=0$.

Furthermore, if we impose the same condition (7) in equation (4), a
straightforward calculation of the sum shows that the temperature is,%

\begin{equation}
T(M)=\frac{T_{H}}{1+\alpha_{1}\left[  \frac{2\hbar\tilde{\Sigma}^{2}A^{4}%
}{16M^{2}+\hbar\tilde{\Sigma}^{2}A^{4}}+\ln(\frac{16M^{2}}{16M^{2}+\hbar
\tilde{\Sigma}^{2}A^{4}})\right]  }.
\end{equation}

It is obvious that removing the quantum corrections i.e., $\alpha_{1}%
=0,$\ leads $T$ to the semi-classical result, $T_{H}$. Significantly, one can
easily verify that $T(M\rightarrow0)\rightarrow0$ and $T(M\rightarrow
\infty)\rightarrow T_{H}$. As it can be seen in Fig. 2(a), when $\alpha_{1}%
<0$\ (the LQG case), the temperature does not take negative value, rather it
remains always positive and goes to zero with $M\rightarrow0$. On the other
hand, for $\alpha_{1}>0$\ (the ST case, see Fig. 2(b)), the temperature does
not exhibit well-behaved behavior as obtained in the LQG case. Because it
first diverges for some finite value of $M$, then becomes negative and
approaches zero from below.

As a final remark for this section, our results suggest that the quantum
corrected Hawking radiation of the LDBH should be considered with the LQG term
$\alpha_{1}<0$ in order to avoid from any unphysical thermodynamical behavior.
Because in the LQG case, both plots of $S(M)$ and $T(M)$ have physically
acceptable thermodynamical behaviors and represent the deserved final;
$S,T(M\rightarrow0)\rightarrow0$.

\section{ENTROPY CONSERVATION OF LDBHs IN QUANTUM CORRECTED HAWKING RADIATION}

In the WKB approximation, the tunneling rate for an outgoing positive energy
particle with a field quantum of energy $\omega$, which crosses the horizon
from $r_{in}(M)$\ to $r_{out}(M-\omega),$ is related to the imaginary part of
the particle's action $\operatorname{Im}(I)$ in accordance with%

\begin{equation}
\Gamma\sim e^{-2\operatorname{Im}(I)}.
\end{equation}
Here $\operatorname{Im}(I)$ is equivalent to%

\begin{align}
\operatorname{Im}(I) &  =-\frac{1}{2}\left[  S(M-\omega)-S(M)\right]
,\nonumber\\
&  =-\frac{1}{2}\Delta S,
\end{align}

which was uncovered in \cite{parikh00}. Let us remark that $\Delta S$ is the
change in entropy of a BH. Hence, the relationship between the tunneling rate
and the entropy change satisfies%

\begin{equation}
\Gamma\sim e^{\Delta S},
\end{equation}

By using (8), $\Delta S$ becomes%

\begin{equation}
\Delta S=\frac{1}{T_{H}}\left\{  -\omega+2\alpha_{1}\ln\left[  \left(
\frac{M-\omega}{\hat{Y}(\omega)}\right)  ^{M-\omega}\left(  \frac{M}{\hat
{Y}(0)}\right)  ^{-M}\right]  \right\}  ,
\end{equation}

where%

\begin{equation}
\hat{Y}(\omega)=\sqrt{\left(  M-\omega\right)  ^{2}+\frac{\hbar\tilde{\Sigma
}^{2}A^{4}}{16}},
\end{equation}

After substituting (13) into (12), the tunneling rate with quantum corrections
to all orders in $\hbar$\ is found as%

\begin{equation}
\Gamma(M;\omega)=\exp\left(  -\frac{\omega}{T_{H}}\right)  \left[  \left(
\frac{M-\omega}{\hat{Y}(\omega)}\right)  ^{M-\omega}\left(  \frac{M}{\hat
{Y}(0)}\right)  ^{-M}\right]  ^{\frac{2\alpha_{1}}{T_{H}}},
\end{equation}

In this expression, the term $\exp\left(  -\frac{\omega}{T_{H}}\right)
$\ arises due to the back reaction effects. The other term to the power
$\frac{2\alpha_{1}}{T_{H}}$\ represents the quantum corrections to all orders
in $\hbar$, and significantly it gives cause for a degeneracy in the pure
thermal radiation. In the absence of the quantum corrections ($\alpha_{1}=0$:
the semi-classical case) the radiation of the LDBH is pure thermal since the
rate (15) reduces to $e^{\frac{-\omega}{T_{H}}}$. The latter case was studied
in detail by \cite{pasa09} in which it was stated that the Hawking radiation
of the LDBH leads to the information loss paradox. The essential annoyance in
the pure thermal radiation is that it never allows the information transfer,
which can be possible with the correlations of the outgoing radiation. So it
is prerequisite to keep the quantum corrections in the tunneling rate (15)
when the agenda is about obtaining a spectrum which is not pure thermal, and
accordingly the correlations of the emitted quanta from the LDBH. In general,
the statistical correlation between two successive emissions is given by
\cite{zhang09,chen09}%

\begin{equation}
\chi(\omega_{1}+\omega_{2};\omega_{1},\omega_{2})=\ln\left[  \frac
{\Gamma\left(  M;\omega_{1}+\omega_{2}\right)  }{\Gamma\left(  M;\omega
_{1}\right)  \Gamma\left(  M;\omega_{2}\right)  }\right]  ,
\end{equation}

and from (15) and (16), one obtains the statistical correlation as%

\begin{equation}
\chi(\omega_{1}+\omega_{2};\omega_{1},\omega_{2})=\frac{2\alpha_{1}}{T_{H}}%
\ln\left[  \frac{\left(  \frac{M-\omega_{1}-\omega_{2}}{\hat{Y}(\omega
_{1}+\omega_{2})}\right)  ^{M-\omega_{1}-\omega_{2}}}{\left(  \frac
{M-\omega_{1}}{\hat{Y}(\omega_{1})}\right)  ^{M-\omega_{1}}\left(
\frac{M-\omega_{2}}{\hat{Y}(\omega_{2})}\right)  ^{M-\omega_{2}}}\right]
\left(  \frac{M}{\hat{Y}(0)}\right)  ^{M},
\end{equation}

This result shows that successive emissions are statistically dependent if and
only if the quantum correction parameter $\alpha_{1}$ is non-zero. Since the
amount of correlation is precisely equal to mutual information between two
sequentially emitted quanta, one can deduce that the statistical correlation
enables the information leakage from the LDBH during its evaporation process.

Now, one can assume that the quasilocal mass of a LDBH is a combination

on of $n$-particles with energies (masses) $\omega_{1},\omega_{2}%
,...\omega_{n}$, $M=%
{\displaystyle\sum\limits_{j=1}^{n}}
\omega_{j}$ in which $\omega_{j}$\ is the energy of the $j^{th}$ emitted field
quanta (particle). Namely, the whole radiation process constitutes of
successively emitted quanta ($\omega_{1},\omega_{2},...\omega_{n}$) from the
BH, so that the LDBH loses its mass $M$ during its evaporation, and at the
final state of the evaporation we find $S,T(M\rightarrow0)\rightarrow0$.

The probability of a radiation composed of correlated quanta is given by the
following product of the tunneling rates \cite{zhang09,zhang11}%

\begin{equation}
\text{P}_{\text{rad}}=\Gamma(M;\omega_{1})\times\Gamma(M-\omega_{1};\omega
_{2})\times....\times\Gamma(M-%
{\displaystyle\sum\limits_{j=1}^{n-1}}
\omega_{j};\omega_{n}),
\end{equation}

where the probability of emission of each radiation of energy $\omega_{j}$ is
given by%

\[
\Gamma(M;\omega_{1})=\exp\left(  -\frac{\omega_{1}}{T_{H}}\right)  \left\{
\left[  \frac{M-\omega_{1}}{Y(\omega_{1})}\right]  ^{M-\omega_{1}}\left[
\frac{M}{\hat{Y}(0)}\right]  ^{-M}\right\}  ^{\frac{2\alpha_{1}}{T_{H}}},
\]

\[
\Gamma(M-\omega_{1};\omega_{2})=\exp\left(  -\frac{\omega_{2}}{T_{H}}\right)
\left\{  \left[  \frac{M-\omega_{1}-\omega_{2}}{Y(\omega_{2})}\right]
^{M-\omega_{1}-\omega_{2}}\left[  \frac{M-\omega_{1}}{Y(\omega_{1})}\right]
^{-\left(  M-\omega_{1}\right)  }\right\}  ^{\frac{2\alpha_{1}}{T_{H}}},
\]

\bigskip%
\begin{equation}
......,
\end{equation}

\[
\Gamma(M-%
{\displaystyle\sum\limits_{j=1}^{n-1}}
\omega_{j};\omega_{n})=\exp\left(  -\frac{\omega_{n}}{T_{H}}\right)  \left\{
\left[  \frac{M-%
{\displaystyle\sum\limits_{j=1}^{n}}
\omega_{j}}{Y(\omega_{n})}\right]  ^{M-%
{\displaystyle\sum\limits_{j=1}^{n}}
\omega_{j}}\left[  \frac{M-%
{\displaystyle\sum\limits_{j=1}^{n-1}}
\omega_{j}}{Y(\omega_{n-1})}\right]  ^{-\left(  M-%
{\displaystyle\sum\limits_{j=1}^{n-1}}
\omega_{j}\right)  }\right\}  ^{\frac{2\alpha_{1}}{T_{H}}},
\]

\[
=\exp\left(  -\frac{\omega_{n}}{T_{H}}\right)  \left[  \frac{\omega_{n}%
}{Y(\omega_{n-1})}\right]  ^{-\frac{2\alpha_{1}}{T_{H}}\omega_{n}},
\]

in which%

\begin{equation}
Y(\omega_{k})=\sqrt{\left(  M-%
{\displaystyle\sum\limits_{j=1}^{k}}
\omega_{j}\right)  ^{2}+\frac{\hbar\tilde{\Sigma}^{2}A^{4}}{16}},
\end{equation}

Here, $\Gamma(M-\omega_{1}-\omega_{2}-....-\omega_{j-1};\omega_{j})$ is the
conditional probability of an emission with energy $\omega_{j}$\ following the
emission before the energy $\omega_{1}+\omega_{2}+....+\omega_{j-1}$.

We can now substitute (19) into (18), and calculate the total probability for
the whole radiation, which turns out to be%

\begin{equation}
\text{P}_{\text{rad}}=\exp\left(  -\frac{M}{T_{H}}\right)  \left(  \frac
{M}{\hat{Y}(0)}\right)  ^{-\frac{2\alpha_{1}M}{T_{H}}},
\end{equation}

According to the statistical mechanics, we recall that all microstates are
equally likely for an isolated system. Since the radiation of a BH can be
considered as an isolated system, the number of microstates $\Omega$ in the
system is $1/$P$_{\text{rad}}$. Thus, one calculate the entropy of the
radiation $S_{rad}$ from the Boltzmann's definition as%

\begin{align}
S_{rad}  &  =\ln\left(  \Omega\right)  =\ln\left(  1/\text{P}_{\text{rad}%
}\right)  ,\nonumber\\
&  =\frac{M}{T_{H}}+\frac{2\alpha_{1}M}{T_{H}}\ln(\frac{M}{\hat{Y}%
(0)}),\nonumber\\
&  =\frac{M}{T_{H}}\left[  1+\alpha_{1}\ln(\frac{16M^{2}}{16M^{2}+\hbar
\tilde{\Sigma}^{2}A^{4}})\right]  .
\end{align}

Clearly, the total entropy of the radiation $S_{rad}$ is equal to the entropy
of the initial LDBH $S(M)$ (8). We deduce therefore that the entropy is
conserved -- the entropy of the original LDBH (before radiation, initial
state) is equal to the entropy of the radiation (after radiation, final
state). From the microscopic point of view of the entropy, this result shows
that the number of microstates of initial and after states are same. The
latter remark implies also that under specific conditions it is possible to
save the information during the Hawking radiation of the LDBHs. In this way,
unitarity in quantum mechanics of the Hawking radiation is also restored.

\section{DISCUSSION AND CONCLUSION}

In this article, we have used SVZR's analysis \cite{singleton10,singleton11}
in order to obtain a specific radiation which yields both zero temperature and
entropy for the LDBH when its mass is radiated away, i.e. $S,T(M\rightarrow0)$
$\rightarrow0$. According to this analysis, the complete evaporation of a BH
is thought as a process in which both back reaction effects and quantum
corrections to all orders in $\hbar$ are taken into consideration. For this
purpose, in Sect. 2 we imposed a condition on $\alpha_{j}$'s which are the
parameters of the quantum corrections to all orders in $\hbar$. Unless the
quantum corrections are ignored, the choice of $\alpha_{j}$'s works finely in
the LDBHs to end up with $S,T(M\rightarrow0)$ $\rightarrow0.$

Upon using the specific form of the entropy (8), we derived the tunneling rate
(15) with quantum corrections to all orders in $\hbar$. Then, it is shown that
this rate attributes to the correlations between the emitted quanta. On the
other hand, existence of the correlations of the outgoing radiation allowed us
to make calculations for the entropy conservation. Thus we proved that after a
LDBH is completely exhausted due to its Hawking radiation, the entropy of the
original LDBH is exactly equal to the entropy carried away by the outgoing
radiation. The important aspect of this conservation is that it provides a
possible resolution for the information loss paradox associated with the
LDBHs. Another meaning of this conservation is that the process of the
complete evaporation of the LDBH is unitary in regard to quantum mechanics.
Because, it is precisely shown that the numbers of microstates before and
after the complete evaporation are the same.

When we analyze the Figs. (1) and (2) which are about the scenario of
$S,T(M\rightarrow0)\rightarrow0$ in the quantum corrected Hawking radiation of
the LDBH, it is seen that our specific choice of $\alpha_{j}$'s (7) with
$\alpha_{1}=\frac{1}{2}$ from ST led to unacceptable behavior for the entropy
(8) in which it gets negative values for some $M$ values. In addition to this,
the behavior of the temperature (9) in the ST case is not well-behaved
compared to the LQG case. However, we have no such unphysical thermodynamical
behaviors in the LQG case. So, for the scenario of $S,T(M\rightarrow
0)\rightarrow0$, we conclude that only the quantum correction term $\alpha
_{1}$ coming from the LQG should be taken into consideration.

In conclusion, we show in detail that the scenario of $S,T(M\rightarrow
0)\rightarrow0$ in the quantum corrected Hawking radiation is possible for the
LDBHs. Furthermore, the information is conserved, and unitarity in quantum
mechanics is restored in the process of complete evaporation of the LDBHs. By
employing SVZR's analysis, we also confirm that quantum corrections with the
back reaction effects remain crucial for the information leakage. Therefore,
it should be stressed that the present study is supportive to the novel idea
introduced by SVZR \cite{singleton10}. Finally, we point out that since the
LDBHs are conformally related to the Brans-Dicke BHs \cite{cai97}, SVZR's
analysis might work for those BHs as well.

\section{Figure Captions}

Figure 1: Entropy $S(M)$ as a function of LDBH mass $M$. The relation is
governed by (8). Figs. 1(a) and 1(b) stand for $\alpha_{1}=-\frac{1}{2}$ and
$\alpha_{1}=\frac{1}{2}$, respectively. The two curves correspond to the
semi-classical entropy (dotted curve) and entropy with quantum corrections to
all orders in $\hbar$ (solid curve).

Figure 2: Temperature $T(M)$ as a function of LDBH mass $M$. The relation is
governed by (9). Figs. 2(a) and 2(b) stand for $\alpha_{1}=-\frac{1}{2}$ and
$\alpha_{1}=\frac{1}{2}$, respectively. The two curves correspond to the
semi-classical temperature (dotted curve) and temperature with quantum
corrections to all orders in $\hbar$ (solid curve).

\end{document}